\documentclass[useAMS]{mn2e}
\usepackage{times}

\title[Equilibria of a Self-gravitating, Rotating Disk]{Equilibria of a Self-Gravitating, Rotating Disk Around a
Magnetized Compact Object}
\author[J. Ghanbari and S. Abbassi]{J. Ghanbari$^{1}$\thanks{E-mail:
ghanbari@Ferdowsi.um.ac.ir ;~~~~ abasi@wali.um.ac.ir } and
S. Abbassi$^{1}$\footnotemark[1]\thanks{}\\
$^{1}$Department of Physics, School of sciences,Ferdowsi University of Mashhad, Mashhad, 91775-1436, Iran\\
$^{}$}
\begin{document}
\date{}

\pagerange{\pageref{firstpage}--\pageref{lastpage}} \pubyear{2004}

\maketitle \label{firstpage}

\begin{abstract}
We examine the effect of self-gravity in a rotating thick-disk
equilibrium in the presence of a dipolar magnetic field. In the
first part, we find a self-similar solution for
non-self-gravitating disks. The solution that we have found shows
that the pressure and density equilibrium profiles are strongly
modified by a self-consistent toroidal magnetic field. We
introduce 3 dimensionless variables $C_B$, $C_c$, $C_t$ that
indicate the relative importance of toroidal component of magnetic
field ($C_B$), centrifugal ($C_c$) and thermal ($C_t$) energy with
respect to the gravitational potential energy of the central
object. We study the effect of each of them on the structure of
the disk. In the second part, we investigate the effect of
self-gravity on the these disks; thus we introduce another
dimensionless variable ($C_g$) that shows the importance of
self-gravity. We find a self-similar solution for the equations of
the system. Our solution shows that the structure of the disk is
modified by the self-gravitation of the disk, the magnetic field
of the central object, and the azimuthal velocity of the gas disk.
We find that self-gravity and magnetism from the central object
can change the thickness and the shape of the disk. We show that
as the effect of self-gravity increases the disk becomes thinner.
We also show that for different values of the star's magnetic
field and of the disk's azimuthal velocity, the disk's shape and
its density and pressure profiles are strongly modified.

\end{abstract}


\section{INTRODUCTION}
The theory of accretion disks, motivated in a large degree by
their occurrence in some binary systems, particularly cataclysmic
variables, has been most fully developed for the thin Keplerian
disks \cite {pringle1}. Based on their geometric shapes, accretion
disks are generally divided into two distinct classes, thin disks
and thick disks. The theory of thin accretion disks (Shakura \&
Sanyev 1973) is well understood whereas there is no universally
accepted model for thick accretion disks. The current interest in
this theory is due to the possibility that thick disks may be
relevant to the understanding of central power sources in radio
galaxies and quasars. Observational evidence suggests that in the
center of many galaxies, matter is somehow ejected to large
distances and gives off high energy radiation as it interacts with
the external medium. There are other theoretical reasons for
pursuing the study of thick disks. In the theory of thin disks,
radial pressure gradients are neglected and vertical pressure
balance is solved separately.It was shown that this approximation
is valid as long as the disk is geometrically thin. This condition
may be violated in the innermost region of accretion disks around
stellar black holes and neutron stars. The study of thick disks
provides a better theoretical understanding of thin disks as a
limiting case and enables us to deal with intermediate situations
\cite{Frank}

When considering the formation process of astrophysical objects,
such as galaxies and stars, the most crucial factor is
self-gravity. In the standard thin accretion disk model, the
effect of self-gravity is neglected, and only pressure supports
the vertical structure. By contrast, the theory of
self-gravitating accretion disks, is less developed. Early
numerical work of self gravitating accretion disks began with
N-body modelling (Cassen \& Moosman 1991 ; Tomley, Cassen \&
Steinman-Cameron 1991). The time evolution of non-self-gravitating
viscose disk has long been studied, and now we have a good theory
describing its steady structure and its basic time-dependent
behavior \cite {shakura}; \cite {pringle}. But with the added
assumption of self-gravity in the disk, it is not easy to follow
its dynamical evolution, mainly because the basic equations for
the disks are highly nonlinear \cite {paz}; \cite {fuk}. To solve
the nonlinear equations of self-gravitating disks, the technique
of self-similar analysis is sometimes useful. Several classes of
self-similar solutions were known previously, but all of them
considered a disk in a fixed, external potential. Self-similar
behavior provides an important class of solutions to the
self-gravitating fluid equations. On the one hand, many physical
problems often attained self-similar limits for a wide range of
initial conditions. On the other hand, the self-similar properties
allows us to investigate properties of solutions in arbitrary
details, without any of the associated difficulties of numerical
hydrodynamics.

Pen (1994) presented a general classification of self-similar
self-gravitating fluids. Fukue \& Sakamoto (1992) also analyzed
the vertical structure of self-gravitating disks, but it is
impossible to compare their models with realistic disks, because
they computed the vertical structure using the thin-disk
approximations for polytropes. Finally, using the numerically
method proposed by Hachisu (1986) and Hachisu et al. (1987),
Woodward et al. (1992) developed a code to investigate the
interaction between a disk and its central object. However, they
only considered polytropic disks. Hashimoto, Eriguchi \& Muller
(1995) presented a two-dimensional equilibrium model for
self-gravitating Keplerian disks. They showed that the shape of
the disk (or disk thickness) in flounced by the rotational law and
the ratio of the disk mass to the mass of the central star. Bodo
\& Curir (1992)computed the equilibrium structure of a
self-gravitating thick accretion disk by an iterative procedure
which produced a final density distribution in equilibrium with
the potential coming from it. They showed that the geometrical
size and shape the disks influenced by self-gravity of the disk.

Accretion disks, containing magnetic fields, have been the subject
of intense study in recent years. The role of the magnetic field
in the equilibrium of accretion disks has been investigated by
some authors \cite{blandford2}; \cite{pringle2} for the thin disk
models , an ideal magneto hydrodynamics (MHD) equilibrium with
azimuthal velocity and poloidal magnetic field has been analyzed
\cite {Lovelace}; \cite {mobarry}. Using numerical methods, these
authors found that the magnetic field may change the shape and
angular momentum distribution in the disk. Thick disk
configurations with a poloidal magnetic field has been studied by
\cite {tri}, in the MHD framework \cite {das}. They investigated
the equilibrium structure of thick disks and their stability in
the presence of a dipolar magnetic field due to a non-rotating
central object. Their solution shows that the pressure and the
density equilibrium profiles are strongly modified by a toroidal
magnetic field, resulting from the interaction between the
permanent dipolar magnetic field and the inertia of the gas disk.
In a magnetized disk, the inertia of the gas is expected to bend
the magnetic field lines backwards, creating a toroidal component,
which in turn may collimate a hydrodynamic outflow over long
distances, forming jets. They assumed that the disk is
non-accreting, stationary, axisymmetric, non-viscous, magnetized
and that it is in equilibrium around a compact object, with only
an azimuthal motion $V_{\phi}$.

We are interested in analyzing the role of self-gravity in thick
disk equilibrium in the presence of the dipolar magnetic field of
a central star. The outline of this paper is as follows: the
general formalism of the problem is discussed in section 2, a
self-similar solution of the equilibrium of non-self-gravitating
accretion disks in the presence of a dipolar magnetic field is
constructed in section 3, a self-similar solution of
self-gravitating magnetized accretion disks is constructed in
section 4 and a summary of the main ideas is given in section5.

\section{GENERAL FORMALISM}

As stated in the introduction, we are interested in analyzing the
role of self-gravity in a thick disk equilibrium in the presence
of the dipolar magnetic field of a central star. For simplicity,
we ignore the influence of energy dissipation. We consider the
disk as a non-accreting MHD flow around a magnetized , non
rotating, central compact object. The disk is assumed to be
stationary and axi-symmetric. We use a spherical polar, inertial,
coordinate system, $(r, \theta, \phi)$, with origin fixed on the
central object. The basic equations are two components of the
euler equation in $(r, \theta)$ direction and poison's equation:
\begin{equation}
\frac{\partial \Psi}{\partial r}=-\frac{1}{4\pi
r}\frac{B_{\phi}}{\rho}\frac{\partial}{\partial r}(r
B_{\phi})-\frac{GM}{r^2}-\frac{1}{\rho}\frac{\partial p}{\partial
r}+\frac{V_{\phi}^{2}}{r},
\end{equation}
\begin{equation}
\frac{\partial \Psi}{\partial \theta}=-\frac{1}{4\pi  \sin
\theta}\frac{B_{\phi}}{\rho}\frac{\partial}{\partial
\theta}(B_{\phi} \sin \theta)-\frac{1}{\rho}\frac{\partial
p}{\partial \theta}+V_{\phi}^{2} \cot \theta,
\end{equation}
\begin{equation}
\frac{1}{r^{2}}\frac{\partial}{\partial
r}(r^{2}\frac{\partial\Psi}{\partial r})+\frac{1}{r^{2} \sin
\theta}\frac{\partial}{\partial \theta}(\sin \theta \frac{\partial
\Psi}{\partial \theta})=4 \pi G\rho,
\end{equation}
where $\rho$, $p$, $V_{\phi}$, $B_{\phi}$ and $\Psi$ denote the
gas density, pressure, toroidal velocity component, toroidal
magnetic field, and gravitational potential respectively. Also G
and M are the gravitational constant and the mass of central star.
Solution of equations (1-3), in general, is a difficult task.
Therefore, to simplify the problem, we impose two constraints. We
adopt for $B_\phi$ and $V_\phi$ the same form as that given in (
Banerjee , Bhatt \& Das; 1995). They assume that the magnetic
field of the central star is dipolar:
\begin{equation}
B_r=2               B_0(\frac{R}{r})^3              \cos\theta,
\end{equation}
\begin{equation}
B_\theta=B_0(\frac{R}{r})^3                         \sin\theta,
\end{equation}
where $B_0$ is the magnetic field strength on the surface of the
central star near the pole and $R$ is its radius. They showed that
the dipolar field with the azimuthal motion of the gas disk, could
make a toroidal component of the magnetic field, that is:
\begin{equation}
B_\phi=B_1(\frac{r}{R})^{(k-\frac{3}{2})} \sin^{-2k} \theta,
\end{equation}
where $B_1$ is an arbitrary constant with the dimension of a
magnetic field strength, and $k\leq\frac{3}{2}$ is a real
constant. The general solution for $V_\phi$ is:
\begin{equation}
V_\phi=V_0(\frac{r}{R})^{(n+\frac{3}{2})} \sin^{-2n} \theta,
\end{equation}
where $R$ denotes the radius of the star. $V_0$  is a constant
with the dimension of a velocity and n is a real constant.

Now, we try to solve our equations with these constraints. At
first we introduce $x$ according to :
\begin{equation}
x=\frac{r}{R_d},~~~~~~~~~~~\alpha=\frac{R_d}{R},
\end{equation}
where $R$ and $R_d$ are the star and the disk radiuses
respectively. Now we rewrite the Eqs.(6)-(7) as:
\begin{equation}
B_\phi=B_1\alpha^{k-\frac{3}{2}}~ x^{k-\frac{3}{2}}~\sin^{-2k}
\theta,
\end{equation}
\begin{equation}
V_\phi=V_0\alpha^{n+\frac{3}{2}}~ x^{n+\frac{3}{2}}~\sin^{-2n}
\theta,
\end{equation}
where $B_1$,$V_0$ are the strength of the toroidal component of
the magnetic field and the rotational velocity on the surface of
the star respectively. We can see clearly that the structure of
the magnetic field of the central star can be modified by a
rotational velocity of the gas disk.

\subsection{MAGNETIC FIELD CONFIGURATION}
In this subsection we want to study the magnetic field of central
stars in the presence of a rotating gas disk. In a magnetized
disk, magnetic field lines are deformed by the rotating gas. In
this case, the gas is expected to be tied to the magnetic field
lines and its inertia causes them to be bent backward and create a
toroidal component.

To study the magnetic field configuration within the disk we will
look at the magnetic field lines, which satisfy the following
equation:
\begin{equation}
\frac{dr}{B_r}=\frac{rd\theta}{B_{\theta}}=\frac{r\sin\theta
d\phi}{B_{\phi}},
\end{equation}

In order to visualize the field line configurations, we now choose
$k =-\frac{3}{2}$ in Eq.6. It is useful to express the results in
a Cartesian frame through the usual relations ($X=r \sin\theta
\cos\phi, Y=r\sin\theta\sin\phi, Z=r\cos \theta $). If we apply
these transformations to the force equation we can obtain the
corresponding parametric equations that generate the magnetic
field configurations:
\begin{equation}
X=r_{in}\cos[\phi_0-\beta(\frac{r_{in}}{R})^3\cos\theta]\sin^3\theta,
\end{equation}
\begin{equation}
Y=r_{in}\sin[\phi_0-\beta(\frac{r_{in}}{R})^3\cos\theta]\sin^3\theta,
\end{equation}
\begin{equation}
Z=r_{in}\sin^2\theta\cos\theta,
\end{equation}
where $\Phi_0$ is a constant of integration (because we consider
axi-symmetry solutions, we can set it to zero without any loss of
generality), $\beta =\frac{B_1}{B_0}$ and $r_{in}$ is the disk
inner radius. From the above equations we can see that the
azimuthal component of the magnetic field can affect magnetic
field line configurations in the disk. In Fig(1) we plot several
2-D field line configurations. We can see that when the toroidal
component of the magnetic field is larger, the deformation of the
magnetic field line is more obvious. In Fig.2 we plot several 3-D
magnetic field configurations with the same parameters.
\input{epsf}
\epsfxsize=3.5in \epsfysize=3.3in
\begin{figure}
\centerline{\epsffile{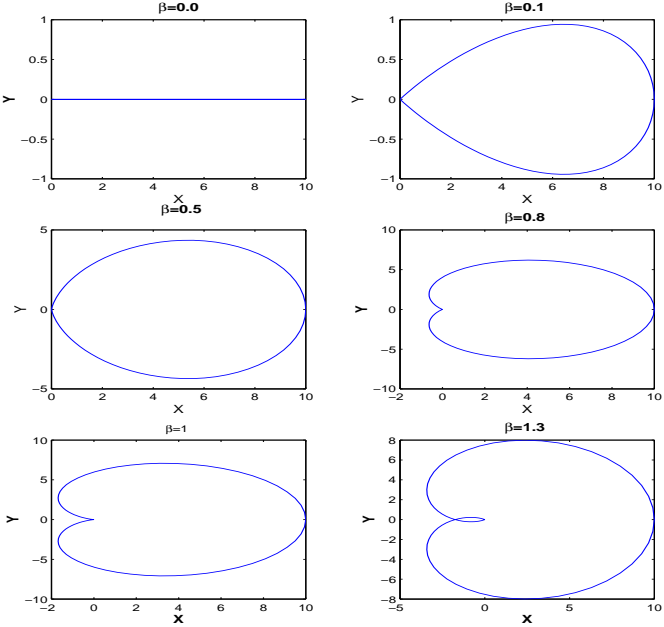}}

 \caption{2D Magnetic field configurations in a X-Y plane as a function of $\theta$,
 which runs from $\theta_{min}=0$ to $\theta_{max}=\pi$
 ,at $r=r_{in}=10$ and R=7 for different value of $\beta=\frac{B_1}{B_0}$ }
 \label{sample-figure}
\end{figure}
\input{epsf}
\epsfxsize=3.5in \epsfysize=3.3in
\begin{figure}
\centerline{\epsffile{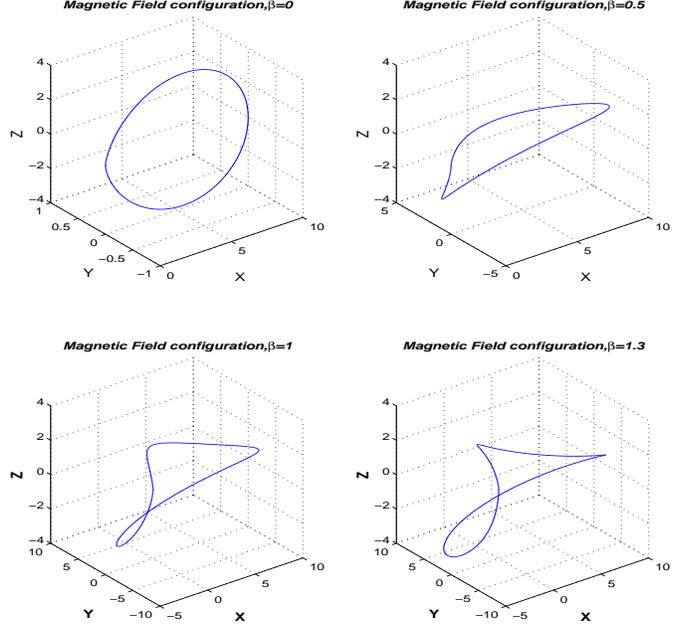}}

 \caption{3D Magnetic field configurations as a function of $\theta$,
 which runs from $\theta_{min}=0$ to $\theta_{max}=\pi$
 ,at $r=r_{in}=10$ and R=7 for different value of $\beta=\frac{B_1}{B_0}$ }
 \label{sample-figure}
\end{figure}

And now we come back to Eqs.(1-3) and introduce other
dimensionless variables for the gravitational potential, the
density and the pressure of the disk's material:
\begin{equation}
S=\frac{\Psi}{\Psi_0},~~~~~~\Sigma=\frac{\rho}{\rho_0},~~~~~~\Lambda=\frac{p}{p_0},
\end{equation}
where $\Psi_0$ , $\rho_0$ and $P_0$ are defined as:
\begin{equation}
\Psi_0=\frac{GM}{R_d}
\end{equation}
\begin{equation}
\rho_0=\rho (r=R_d, \theta=\frac{\pi}{2}),
\end{equation}
\begin{equation}
p_0=p(r=R_d, \theta=\frac{\pi}{2}),
\end{equation}

By inserting these dimensionless parameters into Eqs.(1-2), and by
performing some simplifications, we obtain:
\begin{displaymath}
\frac{\partial S}{\partial x}=-C_B\frac{x^{2k-4}}{\Sigma}
(k-\frac{1}{2})\sin^{-4k}\theta +C_cx^{2n+2}\sin^{-4n}\theta
\end{displaymath}
\begin{equation}
\\-C_t\frac{1}{\Sigma}\frac{\partial \Lambda}{\partial
x}-\frac{1}{x^2},
\end{equation}
\begin{displaymath}
\frac{\partial S}{\partial \theta}=-C_B\frac{x^{2k-3}}{\Sigma}
(-2k+1)\sin^{-4k+1}\theta \cos\theta
\end{displaymath}
\begin{equation}
\\+C_cx^{2n+3}\sin^{-4n}\theta
\cot\theta-C_t\frac{1}{\Sigma}\frac{\partial \Lambda}{\partial
\theta},
\end{equation}
where $C_B, C_c, C_t$ are constants defined by:
\begin{equation}
C_B=\frac{\frac{B^{2}_1}{4\pi\rho_0}}{\frac{GM}{R_d}}\alpha^{2k-3},
\end{equation}
\begin{equation}
C_c=\frac{V^{2}_0}{\frac{GM}{R_d}}\alpha^{2n+3},
\end{equation}
\begin{equation}
C_t=\frac{\frac{P_0}{\rho_0}}{\frac{GM}{R_d}},
\end{equation}

By introducing three dimensionless parameters indicating the
relative importance of the energy of the toroidal component of the
magnetic field ($C_B$), the centrifugal energy ($C_c$), the
thermal energy ($C_t$) with respect to the gravitational potential
energy of the central object, we can study them separately. If
this simplification is used for Poisson's equation, Eq.3, we find
that:
\begin{equation}
\frac{1}{x^2}\frac{\partial}{\partial x}(x^2\frac{\partial
S}{\partial x})+ \frac{1}{x^2 \sin\theta}\frac{\partial}{\partial
\theta}(\sin\theta\frac{\partial S}{\partial
\theta})=\frac{3M_d}{M}\Sigma,
\end{equation}
where $M_d$ is the mass of the disk out to $R_d$:
\begin{equation}
M_d=\frac{4\pi}{3}R^{3}_d \rho_0
\end{equation}
We may introduce another dimensionless variable $C_g$ that gives
the importance of self-gravity. $C_g$ is the ratio of the disk
mass to the central object mass:
\begin{equation}
C_g=\frac{3M_d}{M}
\end{equation}

We would like to investigate self-similar solutions that describe
the equilibrium structure of the thick disk. Thus we introduce
self-similar solutions for the density and pressure of the gas
disk:
\begin{equation}
\Sigma(r,\theta)=\frac{D(\theta)}{x^{\epsilon}}
\end{equation}
\begin{equation}
\Lambda(r,\theta)=\frac{P(\theta)}{x^\delta}
\end{equation}

Since we do not consider energy transport mechanisms, no specific
form for the equation of state is assumed. As a result of our
model we find that the density and pressure clearly show no
polytropic, isothermal or other simple relation. We show that the
indices in the non-self-gravitating and self-gravitating cases are
completely different and that the mass distribution of the disk
changes due to this fact. Solving Eqs.(19-20) and Eqs.(27-28)
gives $\epsilon$ and $\delta$.

\section{ NON SELF-GRAVITATING SOLUTION}

 In this section we try to derive a self-similar solution of non-self-gravitating
 disks. In this regime we put $S$ equal to zero. If we put Eqs.(27-28) into
 Eqs.(19-20) we obtain:
\begin{displaymath}
-(k-\frac{1}{2})C_B
x^{2k-4+\epsilon}\frac{\sin^{-4k}\theta}{D}+C_c
x^{2n+2}\sin^{-4n}\theta
\end{displaymath}
\begin{equation}
\\+\delta C_t x^{\epsilon-\delta-1}\frac{P}{D}-\frac{1}{x^2}=0
\end{equation}
\begin{displaymath}
(2k-1)C_B x^{2k-3+\epsilon}\frac{\sin^{-4k-1}\theta
\cos\theta}{D}+C_c x^{2n+3}\sin^{-4n}\theta \cot\theta
\end{displaymath}
\begin{equation}
\\-C_t x^{\epsilon-\delta}\frac{1}{D}\frac{dP}{d\theta}=0
\end{equation}

In order to obtain the indices of the self-similar solution, we
require that in each equation the exponent of $x$ be the same for
all the terms. Here we get:
\begin{equation}
2k-4+\epsilon=2n+2=\epsilon-\delta-1=-2
\end{equation}
And we have:
\begin{equation}
\epsilon=2-2k
\end{equation}
\begin{equation}
n=-2
\end{equation}
\begin{equation}
\delta=-2k+3
\end{equation}

These equations state that for getting a physical solution, $k$
must be less than unity. Then by putting the self-similar indices
in the equations and by doing some simplifications, we obtain:
\begin{equation}
-(k-\frac{1}{2})C_B \frac{\sin^{-4k}\theta}{D}+C_c
\sin^{8}\theta+(3-2k)C_t \frac{P}{D}-1=0
\end{equation}
\begin{equation}
(2k-1)C_B \frac{\sin^{-4k-1}\theta \cos\theta}{D}+C_c
\sin^{8}\theta \cot\theta-C_t \frac{1}{D}\frac{dP}{d\theta}=0
\end{equation}

We have two equations with two unknowns. The first equation can be
used to express $P$ as a function of $D$:
\begin{equation}
P(\theta)=\frac{1}{(3-2k)C_t}[(1-C_c\sin^{8}\theta)D(\theta)+(k-\frac{1}{2})C_B\sin^{-4k}\theta]
\end{equation}

\input{epsf}
\epsfxsize=3.5in \epsfysize=3.3in
\begin{figure}
\centerline{\epsffile{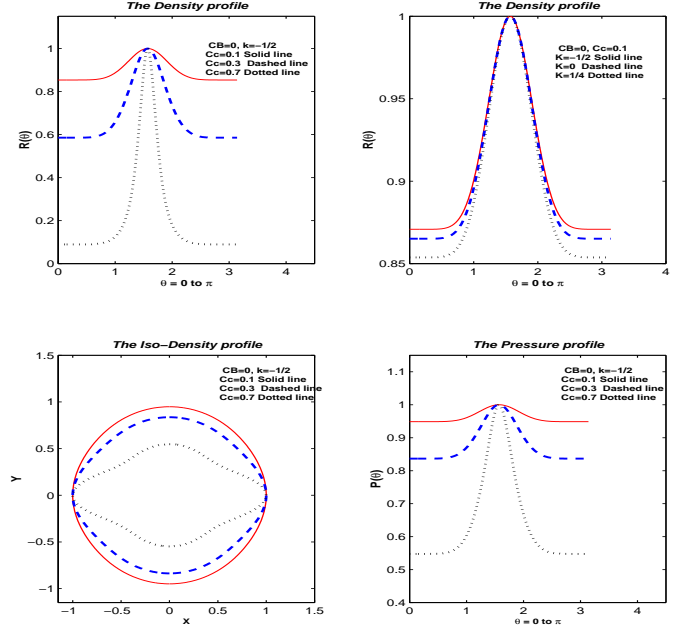}}

 \caption{To the top left:Variation of density along the meridional as a function of $\theta$
 from $\theta_{min}=0$ to $\theta_{max}=\pi$for different value of $C_c$ (rotational velocity), Top right:
 for deferent value of $k$, bottom left: Iso-density contour for different
 value of $C_c$, Bottom right: variation of pressure with different value of $C_c$  }
 \label{sample-figure}
\end{figure}

With our definitions of $\rho_0$ and $P_0$, we have
$D(\theta=\frac{\pi}{2})=1$ and $P(\theta=\frac{\pi}{2})=1$. If we
introduce this boundary condition in Eq.37, $C_t$ can be
calculated:
\begin{equation}
C_t=\frac{1}{(3-2k)}[1-C_c+(k-\frac{1}{2})C_B],
\end{equation}

Now if we introduce Eq.37 into 36, we find:
\begin{equation}
\frac{dD}{d\theta}=g(\theta)D+f(\theta),
\end{equation}
where:
\begin{equation}
g(\theta)=(11-2k)\frac{C_c\sin^{7}\theta\cos\theta}{1-C_c\sin^{8}\theta},
\end{equation}
\begin{equation}
f(\theta)=3(2k-1)C_B\frac{\sin^{-4k-1}\theta\cos\theta}{1-C_c\sin^{8}\theta},
\end{equation}
For this equation we do the following change of variable:
\begin{equation}
D(\theta)=[1-C_c\sin^{8}\theta]^{-\frac{11-2k}{8}}D_{1}(\theta),
\end{equation}
\begin{equation}
\frac{dD_{1}(\theta)}{d\theta}=[1-C_c\sin^{8}\theta]^{\frac{11-2k}{8}}f(\theta),
\end{equation}
or:
\begin{equation}
\frac{dD_{1}(\theta)}{d\theta}=3C_B(2k-1)[1-C_c\sin^{8}\theta]^{\frac{3-2k}{8}}\sin^{-4k-1}\theta
\cos\theta,
\end{equation}

The solution of this equation can be write as:
\begin{equation}
D_{1}(\theta)=3C_B(2k-1)I(\theta)+D_{01},
\end{equation}
where:
\begin{equation}
I(\theta)=\int^{\theta}_{0}[1-C_c\sin^{8}\theta^{'}]^{\frac{3-2k}{8}}\sin^{-4k-1}\theta^{'}
\cos\theta^{'} d\theta^{'};
\end{equation}

Now we come back to our boundary conditions for the density. The
maximum density is at $\theta=\frac{\pi}{2}$ where
$D(\frac{\pi}{2})=1$. Now we can find the value of $D_{01}$ as:
\begin{equation}
D_{1}(\frac{\pi}{2})=[1-C_c\sin^{8}\theta]^{\frac{11-2k}{8}}|_{\theta=\frac{\pi}{2}}D(\frac{\pi}{2})=
[1-C_c]^{\frac{11-2k}{8}};
\end{equation}
Thus:
\begin{equation}
D_{01}=[1-C_c]^{\frac{11-2k}{8}}-3C_B(2k-1)I(\theta=\frac{\pi}{2});
\end{equation}
Hence we can find the density distribution:
\begin{displaymath}
D(\theta)=[(1-C_c)^{\frac{11-2k}{8}}-3C_B(2k-1)(I(\frac{\pi}{2})-I(\theta))]
\end{displaymath}
\begin{equation}
\\\times[1-C_c\sin^{8}\theta]^{-\frac{11-2k}{8}};
\end{equation}

The important result from this equation is that the value of the
density is not zero in the rotation axis. This value depends on
our dimensionless parameters, $C_B$ and $C_c$, where $C_B$
represents the effect of the toroidal component of the magnetic
field and $C_c$ the effect of the rotation velocity.

Fig(3) shows typical examples of the self-similar solution in
equations (37) and (49). In Fig(3), the density profiles, the
iso-density contours and the variations of pressure along the
meridional axes are represented. The density profiles for
different values of $C_c$ (rotational velocity) and $k$ are
displayed. As can be seen, the parameter $C_c$ determines the
overall shape of the matter distribution. By increasing the effect
of the rotational velocity ($C_c$), the shape of the disk changes
to a thinner one, which is due to centrifugal force. In Fig(3), we
also present the iso-density contours for the same value of $C_c$.
As can be seen, the shape of the disk is strongly modified for
different values of $C_c$. We can see that in this case, the
density does not vanish at $\theta= 0, \pi$ (in the rotational
axes). Our solution agrees with \cite{das} solutions for such
systems.

We should mention that the solution we have found is a
mathematical one.  As a result, the solution has a physical
meaning only for parts of the parameter space. In this
self-similar solution we have a singularity in $\theta =0, \pi$
that arises from $\sin\theta$ in the integral of equation (46);
Thus we cannot study the effect of the magnetic field in the
equilibrium structure of the gas disk. That is the nature of the
solution. After we have found the analytical solution for
non-self-gravitating disks, we will try to investigate the effect
of self-gravity in such disks.

\section{SELF-GRAVITATING SOLUTION}
In order to gain a better understanding of the physical situation
of the disk, we include self-gravity in the above equations. The
self-similar technique is utilized once again. If we introduce
Eq.(27-28) in to
 Eq. (19-20) we obtain:
\begin{displaymath}
\frac{\partial S}{\partial
x}=-C_B(k-\frac{1}{2})x^{2k-4+\epsilon}\frac{\sin^{-4k}\theta}{D(\theta)}
+C_cx^{2n+2}\sin^{-4n}\theta
\end{displaymath}
\begin{equation}
\\+\delta C_t
x^{\epsilon-\delta-1}\frac{P(\theta)}{D(\theta)}-\frac{1}{x^2};
\end{equation}
\begin{displaymath}
\frac{\partial S}{\partial
\theta}=-C_B(-2k+1)x^{2k-3+\epsilon}\frac{\sin^{-4k-1}\theta \cos
\theta}{D(\theta)}
\end{displaymath}
\begin{equation}
\\+C_cx^{2n+3}\sin^{-4n}\theta \cot \theta-
C_tx^{\epsilon-\delta}\frac{1}{D(\theta)}\frac{dP(\theta)}{d\theta};
\end{equation}

For finding the self-similar solution we must set equal the terms
that have the same power.  We obtain:
\begin{equation}
\epsilon=\frac{5-2k}{2}
\end{equation}
\begin{equation}
n=\frac{2k-7}{4}
\end{equation}
\begin{equation}
\delta=3-2k
\end{equation}

Now we can identify the radial dependence of the density and
pressure of the gas disk ($\epsilon$ and $\delta$ ) as functions
of $k$. We can insert Eqs.(50-51) into Eq.(24) with use Eq.(27) we
can find:
\begin{equation}
\frac{\partial S}{\partial x}=
H(\theta)x^{\frac{2k-3}{2}}-\frac{1}{x^2},
\end{equation}
\begin{equation}
\frac{\partial S}{\partial \theta}= W(\theta)x^{\frac{2k-1}{2}},
\end{equation}
where we used a $k$ as a free parameter and:
\begin{displaymath}
H(\theta)=[-(k-\frac{1}{2})C_B\frac{\sin^{-4k}\theta}{D(\theta)}
+C_c\sin^{-2k+7}\theta
\end{displaymath}
\begin{equation}
~~~~~~~~~~~~~~~~~~~~~~~~~~~~~~~~~~+(3-2k)C_t\frac{P(\theta)}{D(\theta)}],
\end{equation}
\begin{displaymath}
W(\theta)=[C_B(-2k+1)\frac{\sin^{-4k-1}\theta \cos
\theta}{D(\theta)} +C_c\sin^{-2k+7}\theta \cot \theta
\end{displaymath}
\begin{equation}
~~~~~~~~~~~~~~~~~~~~~~~~~~~~~~~~~~~~~~-C_t\frac{1}{D(\theta)}\frac{dP}{d\theta}],
\end{equation}

Now if we introduce Eqs.(55-56) into Eq. (24) and do some
simplifications, we obtain an ordinary differential equation with
two unknown functions ($D(\theta)$ and $P(\theta)$):
\begin{equation}
(\frac{2k+1}{2})H(\theta)+\frac{1}{\sin
\theta}\frac{d}{d\theta}[\sin \theta
W(\theta)]=\frac{3M_d}{M}D(\theta)
\end{equation}

In order to solve this equation, we need another relation between
$D(\theta)$ and $P(\theta)$. By looking at the definitions of
$H~\&~W$, we can find:
\begin{equation}
\frac{2}{2k-1}\frac{dH}{d\theta}=W(\theta),
\end{equation}

\input{epsf}
\epsfxsize=2.8in \epsfysize=3.3in
\begin{figure}
\centerline{\epsffile{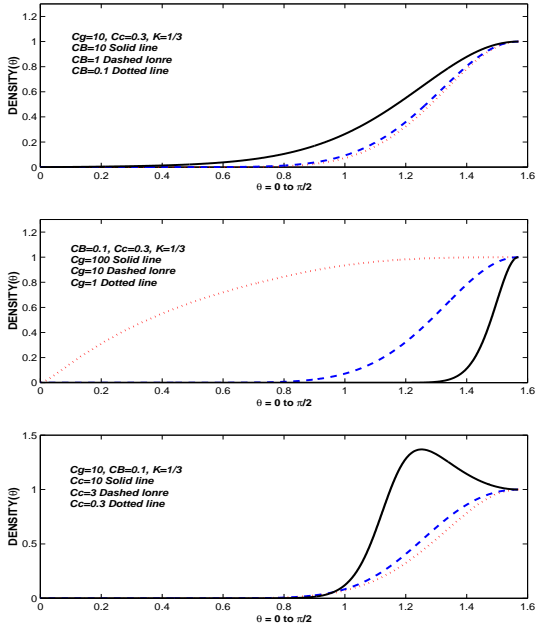}}

 \caption{Variation of the density of the disk for several values of
 the parameters:
 the top one: is for different values of $C_B$ that represents the
 importance of the magnetic field; the middle one: for different
 values
 of $C_g$ that represents the importance of self-gravity; the bottom one: is for
 different values of $C_c$ that represents the importance of the rotational
 velocity  }
 \label{sample-figure}
\end{figure}
\input{epsf}
\epsfxsize=2.8in \epsfysize=3.3in
\begin{figure}
\centerline{\epsffile{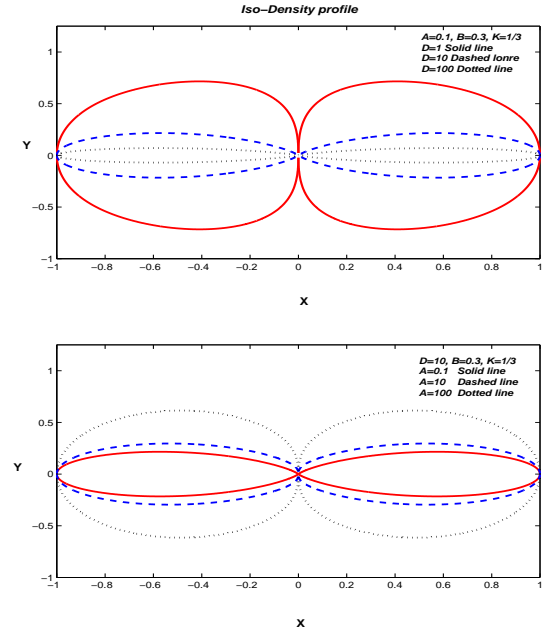}}

 \caption{Iso-density contours of the self-gravitating disks: the top one:
 for different values of $C_g$ that represents the importance of self-gravitation;
 the bottom: for different values of $C_B$
 that represents the importance of the magnetic energy.  }
 \label{sample-figure}
\end{figure}

Now by solving Eqs. (59-60) we can find the angular dependence of
the density and pressure of the gas disk ($D, P$). We linearize
these equations, introduce another dimensionless constant defined
as the ratio of the disk mass to the mass of the central object
($C_g$), and we use $Q=\frac{1}{D}$ instead of $D$ in order to
simplify the equations. Thus with some calculations, we get:
\begin{equation}
\frac{dy_1}{d\theta}=S(\theta,y_1 ,y_2 ,y_3);
\end{equation}
\begin{equation}
\frac{dy_2}{d\theta}=y_3;
\end{equation}
\begin{equation}
\frac{dy_3}{d\theta}=T(\theta,y_1 ,y_2 ,y_3);
\end{equation}
where $y_1=Q=\frac{1}{D}$, $y_2=P$, $y_3=\frac{dP}{d\theta}$,
 and $S$ and $T$ are complex functions of $\theta$ ,
$y_1$, $y_2$, $y_3$. We solve these linear equations, with two
point boundary conditions with the shooting method. Solving these
equations gives the meridional component of the density and
pressure. We used Naryan \& Yi (1995) boundary conditions:
\begin{equation}
 \frac{d\rho}{d\theta}\mid_{\theta=0}=\frac{d\rho}{d\theta}\mid_{\theta=\frac{\pi}{2}}=0
\end{equation}
\begin{equation}
  \frac{dp}{d\theta}\mid_{\theta=0}=\frac{dp}{d\theta}\mid_{\theta=\frac{\pi}{2}}=0
\end{equation}

By changing the value of the dimensionless parameters $C_B$,
$C_c$, $C_t$ and $C_g$, we can study the influence of the toroidal
magnetic field, toroidal velocity, thermal energy and self-gravity
of the gas disk on its own equilibrium structure. In Fig( 4-6) we
plot some of our results. By comparing the results in Fig(4-5)
which are calculated for different values of $C_g$, we can see how
the disk is more sensitive to the influence of self-gravity.
Fig(4) shows the variations of the density for different values of
$C_g$ , $C_B$, $C_c$ in meridional direction. Concerning the
geometrical shape of the disk we can see the tendency of the disk
to become thinner when self-gravity plays an important role ($C_g$
is large). And then we can see that a strong magnetic field coming
from the central star can produce a large toroidal component,
which can change the thickness of the disk near the pole. Fig(4)
shows the iso-density contours of the disk for different values of
$C_g$ and $C_B$. In Fig(6) we present the iso-density profile of
the disk for different values of $C_c$ and the iso-pressure
profile of the disk for different values of $C_B$. We can see in
the iso-pressure profile that near the pole there is a
singularity; maybe this big pressure gradient could produce a jet
from the pole! In order to test this effect we should investigate
the time evolution of such systems.  This could be the subject of
future works.

Concerning the ratio of the mass of the disk to the mass of the
central star, Eriguchi \& Muller (1993) extended their code to
include a central object for a thick disk and calculated the
equilibrium structure of the disk which rotates according to the
$j$-const law, where $j$ is the specific angular momentum.
Comparing this with the toroidal star of Hashimito et al. (1993),
where the central object is not included, the relative thickness
of the disk is smaller, which is caused by effects due to gravity
of the central star (see figures in Eriguchi \& Muller 1993). As
for the rotation law, Hashimito et al. (1993) and Eriguchi \&
Muller (1993) obtained rather thick disks, while Hashimito et al.
(1995) found a flat disk shape where the gravitational force of
the central object becomes relatively week compared with the
self-gravity of the disk. Our results agree with the latter one.
We can see in Fig(4-6) that the shape of the disk (or disk
thickness) is influenced by the rotational law and the ratio of
the disk mass to the mass of the central star.

\input{epsf}
\epsfxsize=3.3in \epsfysize=3.3in
\begin{figure}
\centerline{\epsffile{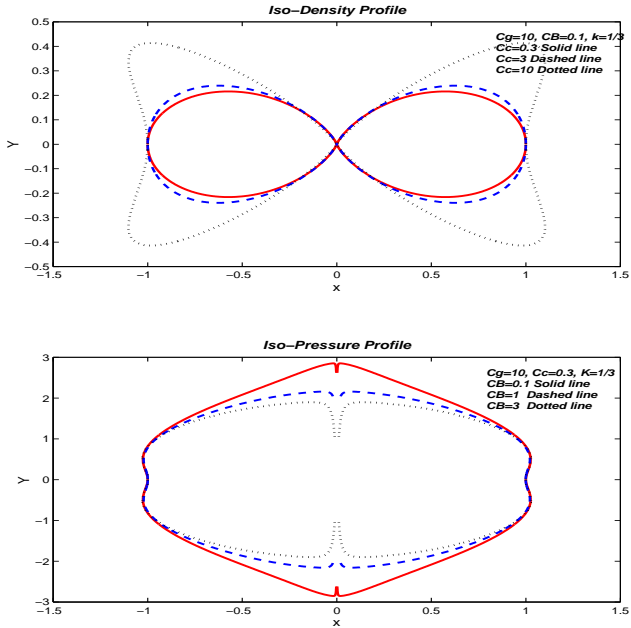}}

 \caption{Iso-density contours of the self-gravitating accretion disks: the top one:
 for different values of $C_c$, which represents the importance of rotational velocity;
 the bottom one: iso-pressure contours of the disk for
 different values of $C_B$, which represents the importance of the magnetic field. }
 \label{sample-figure}
\end{figure}

\section{DISCUSSION AND CONCLUSIONS}

In this paper, we find a global equilibrium of a self-gravitating
disk around a magnetized compact object. We ignore the effects of
energy dissipation and radial flow. We have investigated a
stationary model for a self-gravitating disk influenced by a
strong magnetic field coming from the central object. We ignore
the effects of the local magnetic field of the disk. We consider
that the radial flow is negligible with respect to the azimuthal
motion of the gas in the disk. We find two types of self-similar
solutions for such disks. The first one is the
non-self-gravitating self-similar solution that shows that the
shape of the disk changes when the physical parameters are
modified. The second one is the self-gravitating self-similar
solution. In this class of solutions we can study the effect of
self-gravity on the shape and thickness of the disk.

In the non-self-gravitating solution we find an analytical
solution for some parts of the parameter space. This solution is
found with a self-similarity method, and it gives a picture of the
equilibrium structure of the gas material around a magnetized
compact object. The solution shows that the rotational velocity of
the gas disk can change the equilibrium picture. The important
result that can be inferred from this solution is the
non-vanishing value of the density on the rotation axis. This
value depends on the dimensionless parameters, $C_B$ and $C_c$,
where $C_B$ represents the effect of the magnetic field and $C_c$
the effect of the rotational velocity.

In the self-gravitating solution, we find the vertical
(meridional) dependence of the gas density in the disk Fig(4). The
external dipole field can change the thickness of the disk near
the pole, and the existence of a sharp density gradient in the
$\theta$-direction indicates a rapid decrease of the matter
density away from the equatorial plane. But when self-gravity
plays an important role in the disk equilibrium (by increasing
$C_g$) we can see that the thickness of the disk decreases.

 In Fig(5-6) we show some iso-density and iso-pressure contours
 of the equilibrium configuration of the disk for
different values of $C_B$, $C_c$, $C_g$, $k$. In the iso-pressure
profile we can see a large pressure gradient near the poles, which
may lead to outflows in those regions! This idea would require
further study, such as a dynamical investigation.

We see that the presence of self-gravity in a thick disk, can
change the geometrical shape of the disk and plays an important
role in equilibrium structure of the disk. We also see that the
strength of the magnetic field can change the structure of the
disk near the poles. Finally, we must admit that our model does
not deserve the name ''accretion'' disk, since we did not include
the accretion (mass) flow.

This study, one of the first of its kind, is a search for an
equilibrium structure of a thick disk in the presence of an
external stellar dipole along with a self-consistently generated
toroidal magnetic field $B_{\phi}$. The meridional structure of
the disk is mainly due to the balance of plasma pressure gradient,
magnetic force due to $B_{\phi}$, and centrifugal force. The
existence of an equilibrium structure, in fact, encourages one to
look for generalizations of the analysis to the case when the
radial velocity $V_r\neq 0$, representing accretions. On the other
hand, yet another important aspect to be considered is the
generalization of the newtonian analysis to general relativistic
formalism wherein space-time curvature produced by the strong
gravitational field of the central object would modify the
magnetic fields and introduce new features.

Our disk model might also apply to systems where the central
object is a black hole, a neutron star or an active galactic
nucleus. However, in this paper we modelled the disk under the
assumption of a Newtonian potential. For a system where the
central object is a black hole, the pseudo-Newtonian potential
will play an essential role near the central object. The
modification arising from a pseudo-Newtonian potential to the disk
will be investigated in a future paper. Consequently, we cannot
calculate the accretion luminosity. Thus, to compare our results
with other models or observations, it will be necessary to include
mass flow in our model.

We are grateful to the referee for a very careful reading of the
manuscript and for suggestions which help us improve the
presentation of our results. We are grateful Mohsen Shadmehri for
continuous encouragement and useful discussions. We thank Kattia
Ferriere, Daniel Reese and saeed farivar for their useful
comments.

\end{document}